\documentclass[5p,preprint,twocolumn]{elsarticle}

\usepackage{hyperref}
\usepackage{lineno}
\modulolinenumbers[5]

\usepackage{graphicx}
\usepackage{bm}
\hypersetup{colorlinks=false}

\journal{Physics Letters B}

\begin{document}

\begin{frontmatter}

\title{Class of solutions of the Wheeler-DeWitt equation with ordering parameter}


\author[mymainaddress]{H. S. Vieira\corref{mycorrespondingauthor}}
\cortext[mycorrespondingauthor]{Corresponding author}
\ead{horacio.santana.vieira@hotmail.com}

\author[secondaryaddress]{V. B. Bezerra}
\ead{valdir@fisica.ufpb.br}

\author[thirdaddress]{C. R. Muniz}
\ead{celio.muniz@uece.br}

\author[fourthaddress]{M. S. Cunha}
\ead{marcony.cunha@uece.br}

\author[fifthaddress]{H. R. Christiansen}
\ead{hugo.christiansen@ifce.edu.br}

\address[mymainaddress]{Instituto de F\'{i}sica de S\~{a}o Carlos, Universidade de S\~{a}o Paulo, Caixa Postal 369, CEP 13560-970, S\~{a}o Carlos, S\~{a}o Paulo, Brazil}
\address[secondaryaddress]{Departamento de F\'{i}sica, Universidade Federal da Para\'{i}ba, Caixa Postal 5008, CEP 58051-970, Jo\~{a}o Pessoa, PB, Brazil}
\address[thirdaddress]{Grupo de F\'isica Te\'orica (GFT), Universidade Estadual do Cear\'a, Faculdade de Educa\c c\~ao, Ci\^encias e Letras de Iguatu, Iguatu, Cear\'a, Brazil}
\address[fourthaddress]{Grupo de F\'isica Te\'orica (GFT), Centro de Ci\^encias e Tecnologia, Universidade Estadual do Cear\'a, CEP 60714-903, Fortaleza, Cear\'a, Brazil}
\address[fifthaddress]{Grupo de F\'isica Te\'orica (GFT), Instituto Federal de Ci\^encias, Educa\c c\~{a}o e Tecnologia, CEP 62042-030, Cear\'a, Brazil}

\begin{abstract}
In this letter, we discuss the Wheeler-DeWitt equation with an ordering parameter in the Friedmann-Robertson-Walker universe. The solutions when the universe was very small and at the end of the expansion are obtained in terms of Bessel and Heun functions, respectively. We also obtain a boundary condition which should be satisfied by the ordering parameter, namely, $0 \leq p \leq 2$. We investigate the minimum value of the scale factor with respect to the maximum value of the probability density.
\end{abstract}



\end{frontmatter}


%
%
\section{Introduction}
In order to formulate a quantum cosmology theory for the evolution of the universe, DeWitt \cite{PhysRev.160.1113} and Wheeler \cite{Wheeler:1968} proposed an equation inspired on the Hamilton-Jacobi approach to General Relativity \cite{NuovoCimento.26.53}. In their formulation, the wave function of the universe can be obtained and as a consequence, it is possible, in principle, to discuss the quantum properties of the spacetime following DeWitt and Wheeler ideas.

This line of research, called quantum relativistic cosmology, has inspired a lot of investigations such as the tunneling wave function of the universe \cite{PhysRevD.99.066010}, the interaction with another universe and effects on the cosmic microwave background \cite{JCAP.1902.057}, as well as some aspects concerning the consistence of the Wheeler-DeWitt equation in modified theories of gravity \cite{JCAP.1812.032}, among others.

In the middle 1980's, Vilenkin \cite{PhysRevD.33.3560} analyzed the problem of boundary conditions in quantum cosmology. In his work, the Wheeler-DeWitt equation plays the role of the Schr\"{o}dinger equation for the wave function of the universe. The evolution from a universe with ``nothing'' to a large scale one was his great achievement. Based in this approach, He \textit{et al}. \cite{PhysLettB.748.361} studied the dynamical interpretation of the wave function of the universe by considering the Wheeler-DeWitt equation with an ordering parameter in the Friedmann-Robertson-Walker universe with open geometry.

Recently, we found a class of solutions of the Wheeler-DeWitt equation in a homogeneous and isotropic universe for the spatially closed, flat, and open geometries of the Friedmann-Robertson-Walker universe filled with different forms of energy, namely, matter, radiation, and vacuum \cite{PhysRevD.94.023511}. Soon after, we have generalized our analysis and then studied the dynamical interpretation of the Wheeler-DeWitt equation by taking into account all different kinds of energy \cite{arXiv:1904.10864v1}.

In this work we examine the Wheeler-DeWitt equation with an ordering parameter in a universe filled with several kinds of energy, namely, vacuum, matter, radiation, string network, and quintessence. Furthermore, we do not fix a geometry for the spacetime, that is, the universe can be spatially closed, flat, or open.

This Letter is organized as follows. In Section \ref{Wheeler-DeWitt} we present the Wheeler-DeWitt equation with an ordering parameter in the Friedmann-Robertson-Walker universe. In Section \ref{solutions} we discuss the asymptotic behaviors of the wave function of the universe and obtain a boundary condition which should be satisfied by the ordering parameter. Finally, in Section \ref{Summary} we close the paper with the final remarks.
%
%
\section{Wheeler-DeWitt equation with ordering term}\label{Wheeler-DeWitt}
We want to discuss some features of the quantum relativistic cosmology. In order to do this, let us establish the Wheeler-DeWitt equation in the minisuperspace approximation as follows.

The classical evolution of the universe at different stages, valid for all forms of energy, can be obtained from the following Hamiltonian \cite{PhysRevD.94.023511}
\begin{equation}
H(P_{a},a)=\frac{3 \pi c^{2}}{4G}a^{3}\biggl(\frac{4G^{2}P_{a}^{2}}{9 \pi^{2} c^{4} a^{4}}+\frac{k c^{2}}{a^{2}}-\frac{8 \pi G}{3 c^{2}}\rho\biggr),
\label{eq:Hamiltonian_WDW}
\end{equation}
where $P_{a}$ is the canonically conjugate momentum, $k=-1,0,+1$ is the curvature parameter, and $\rho$ is the energy density. The scale factor, $a$, is such that $0 \leq a < \infty$.

Now, in order to write down the most general form of the Wheeler-DeWitt equation, we need to do the quantization of the momentum, as follows
\begin{equation}
P_{a}^{2} \rightarrow -\frac{\hbar^{2}}{a^{p}}\frac{\partial}{\partial a}\biggl(a^{p}\frac{\partial}{\partial a}\biggr),
\label{eq:momentum}
\end{equation}
and then to impose the constraint $H\Psi(a)=0$, where $\Psi(a)$ is the wave function of the universe which depends only on the scale factor. The parameter $p$ represents the uncertainty in the ordering of the operator factors $a$ and $\partial/\partial a$ \cite{PhysRevD.33.3560}.

Thus, the Wheeler-DeWitt equation with an ordering parameter, which corresponds to the term with the ordering factor $p$, in the Friedmann-Robertson-Walker universe, can be written as
\begin{equation}
\biggl[\frac{d^{2}}{da^{2}}+\frac{p}{a}\frac{d}{da}-\frac{9\pi^{2}c^{4}a^{2}}{4\hbar^{2}G^{2}}\biggl(kc^{2}-\frac{8 \pi G a^{2}}{3c^{2}}\rho\biggr)\biggr]\Psi=0.
\label{eq:WDE_ordering}
\end{equation}
The total energy density $\rho$ can be written as the sum of all kinds of energy, namely, vacuum, matter, radiation, string network and quintessence. It is given by
\begin{equation}
\rho=\sum_{\omega}\rho_{\omega}=\rho_{v}+\rho_{m}+\rho_{r}+\rho_{s}+\rho_{q}.
\label{eq:density_sums}
\end{equation}
The energy density of the vacuum, $\rho_{v}$, is expressed in terms of the cosmological constant as
\begin{equation}
\rho_{v}=\frac{\Lambda c^{4}}{8\pi G}.
\label{eq:WDE_energy_density_vacuum}
\end{equation}
For the other kinds of energy, the energy density $\rho_{\omega}$ is given by
\begin{equation}
\rho_{\omega}=\frac{A_{\omega}}{a^{3(\omega+1)}},
\label{eq:WDE_density}
\end{equation}
where the energy density parameter $A_{\omega}$ is summarized in Table \ref{tab:A_omega_WDW_orering}.

\begin{table}[!ht]
\caption{The energy density parameter $A_{\omega}=\rho_{\omega 0}a_{0}^{3(\omega+1)}$ related to the state parameter $\omega$. Here $\rho_{\omega 0}$ stands for the value of $\rho_{\omega}$ at present time, as well as $a_{0}$ does for $a$.}
\label{tab:A_omega_WDW_orering}
\begin{tabular}{crl}
\hline\noalign{\smallskip}
			Nature of energy  & $\omega$ 				& $A_{\omega}$ \\
\noalign{\smallskip}\hline\noalign{\smallskip}
			matter						& 0 							& $A_{m}=\rho_{m0}a_{0}^{3}$ \\
			radiation 				& $\frac{1}{3}$ 	& $A_{r}=\rho_{r0}a_{0}^{4}$ \\
			string network 			& $-\frac{1}{3}$ 	& $A_{s}=\rho_{s0}a_{0}^{2}$ \\
			quintessence 			& $-\frac{2}{3}$ 	& $A_{q}=\rho_{q0}a_{0}$ \\
\noalign{\smallskip}\hline
\end{tabular}
\end{table}

In what follows, we will obtain the class of solutions of the Wheeler-DeWitt equation given by Eq.~(\ref{eq:WDE_ordering}).
%
%
\section{Class of solutions}\label{solutions}
In order to solve the Wheeler-DeWitt equation with an ordering parameter, let us substitute Eqs.~(\ref{eq:WDE_energy_density_vacuum}) and (\ref{eq:WDE_density}) into Eq.~(\ref{eq:WDE_ordering}) and assume that the wave function can be written as $\Psi(a)=a^{-p/2}y(a)$. Thus, we obtain
\begin{equation}
-\hbar^{2}\frac{d^{2}y(a)}{da^{2}}+V_{eff}(a,p)y(a)=0,
\label{eq:WDE_mov_1}
\end{equation}
where $V_{eff}(a,p)$ is the effective potential given by
\begin{eqnarray}
V_{eff}(a,p) & = & \frac{9\pi^{2}c^{6}k}{4G^{2}}a^{2}-\frac{6\pi^{3}c^{2}}{G}\biggl(A_{r}+A_{m}a\nonumber\\
&& +A_{s}a^{2}+A_{q}a^{3}+\frac{\Lambda c^{4}}{8\pi G}a^{4}\biggr)\nonumber\\
&& +\frac{(p^2-2 p) \hbar ^2}{4}\frac{1}{a^{2}}.
\label{eq:WDW_effective_potential_energy}
\end{eqnarray}

At this point we can compare this effective potential with the one that we obtained in Ref.~\cite{arXiv:1904.10864v1}. In fact, the unique difference is the last term proportional to $1/a^{2}$, which depends on the ordering parameter $p$. This means that the term with the ordering parameter will contribute significantly only in the case when the scale factor goes to zero, that is, in the beginning of the universe.

For the simple minisuperspace model, Hawking \textit{et al.} \cite{NuclPhysB.264.185} proposed that the differential operator in the Wheeler-DeWitt equation should be the Laplace operator and hence $p=1$. On the other hand, He \textit{et al.} \cite{PhysLettB.748.361} analyzed the dynamical interpretation of the wave function of the universe and obtained two boundary conditions which determine the value of the ordering parameter as being $p=-2$.

\begin{figure}[!ht]
		\includegraphics[scale=0.35]{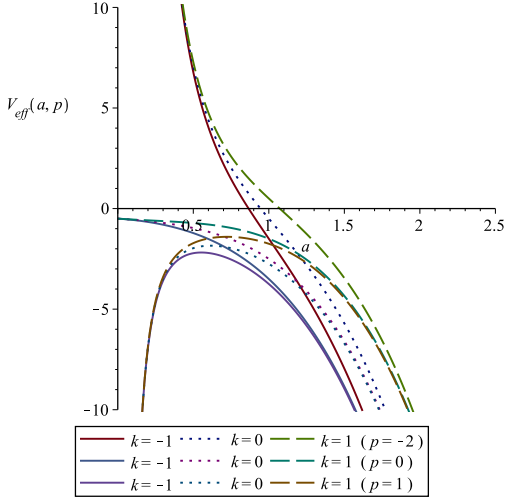}
	\caption{The effective potential $V_{eff}(a,p)$ for $\Lambda > 0$. We focus on the $p=-2,1$ cases and compare with the case where $p=0$.}
	\label{fig:WDW_Fig1}
\end{figure}

\begin{figure}[!ht]
		\includegraphics[scale=0.35]{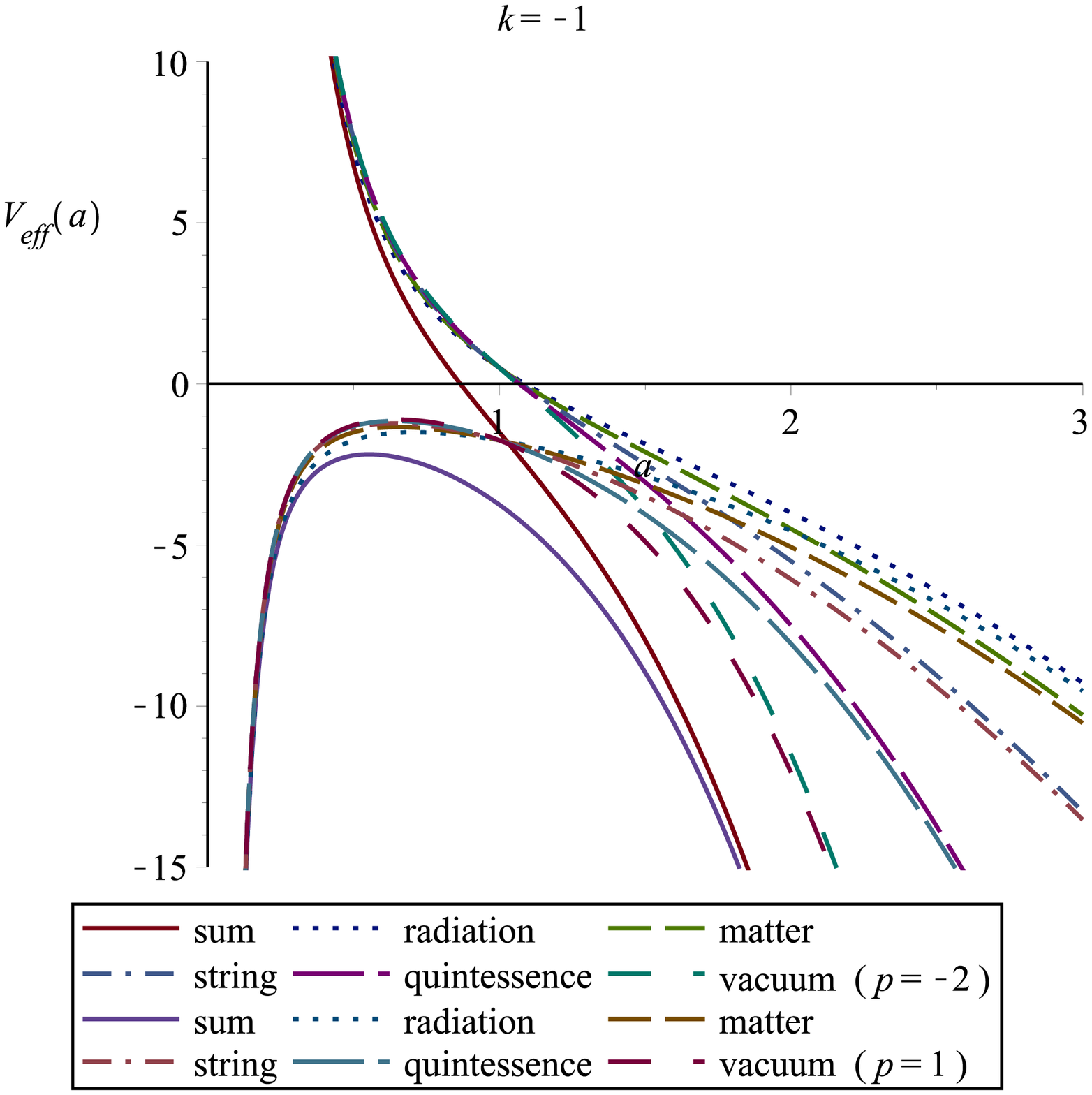}
	\caption{The effective potential $V_{eff}(a,p)$ for $\Lambda > 0$.  We focus on the $k=-1$ case and compare with each kind of energy density $\omega$.}
	\label{fig:WDW_Fig2}
\end{figure}

\begin{figure}[!ht]
		\includegraphics[scale=0.35]{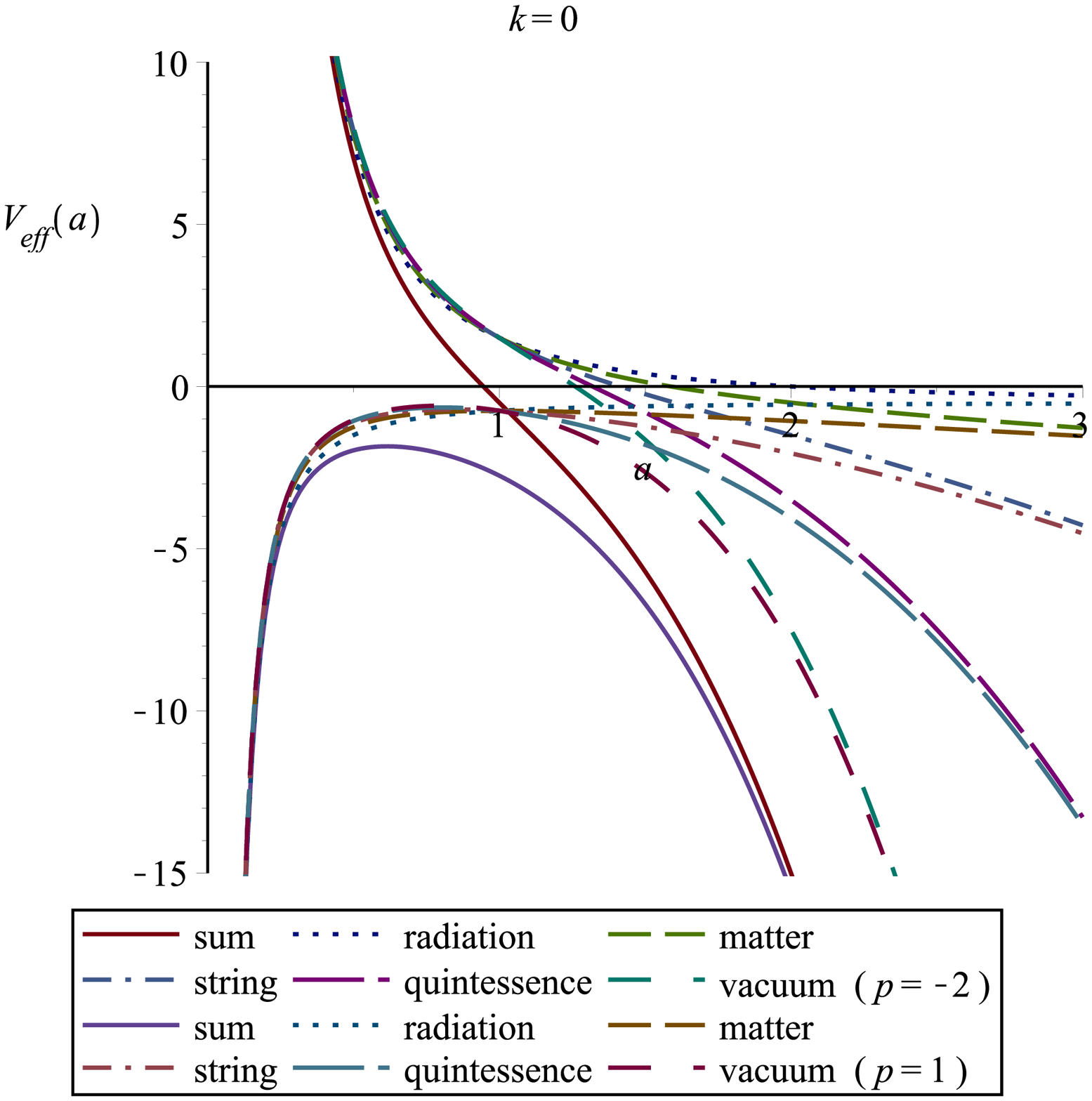}
	\caption{The effective potential $V_{eff}(a,p)$ for $\Lambda > 0$.  We focus on the $k=0$ case and compare with each kind of energy density $\omega$.}
	\label{fig:WDW_Fig3}
\end{figure}

\begin{figure}[!ht]
		\includegraphics[scale=0.35]{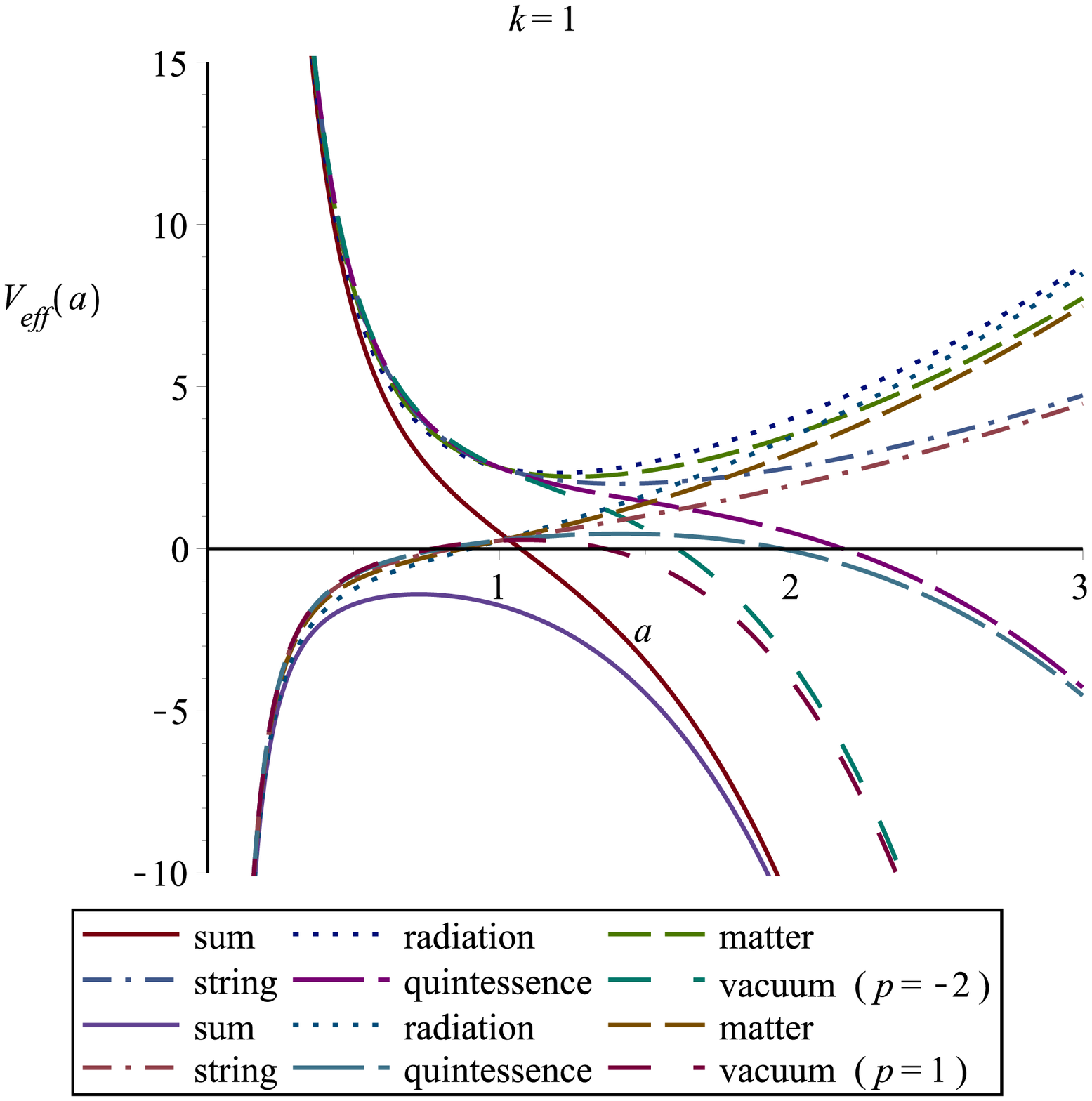}
	\caption{The effective potential $V_{eff}(a,p)$ for $\Lambda > 0$.  We focus on the $k=+1$ case and compare with each kind of energy density $\omega$.}
	\label{fig:WDW_Fig4}
\end{figure}

\begin{figure}[!ht]
		\includegraphics[scale=0.35]{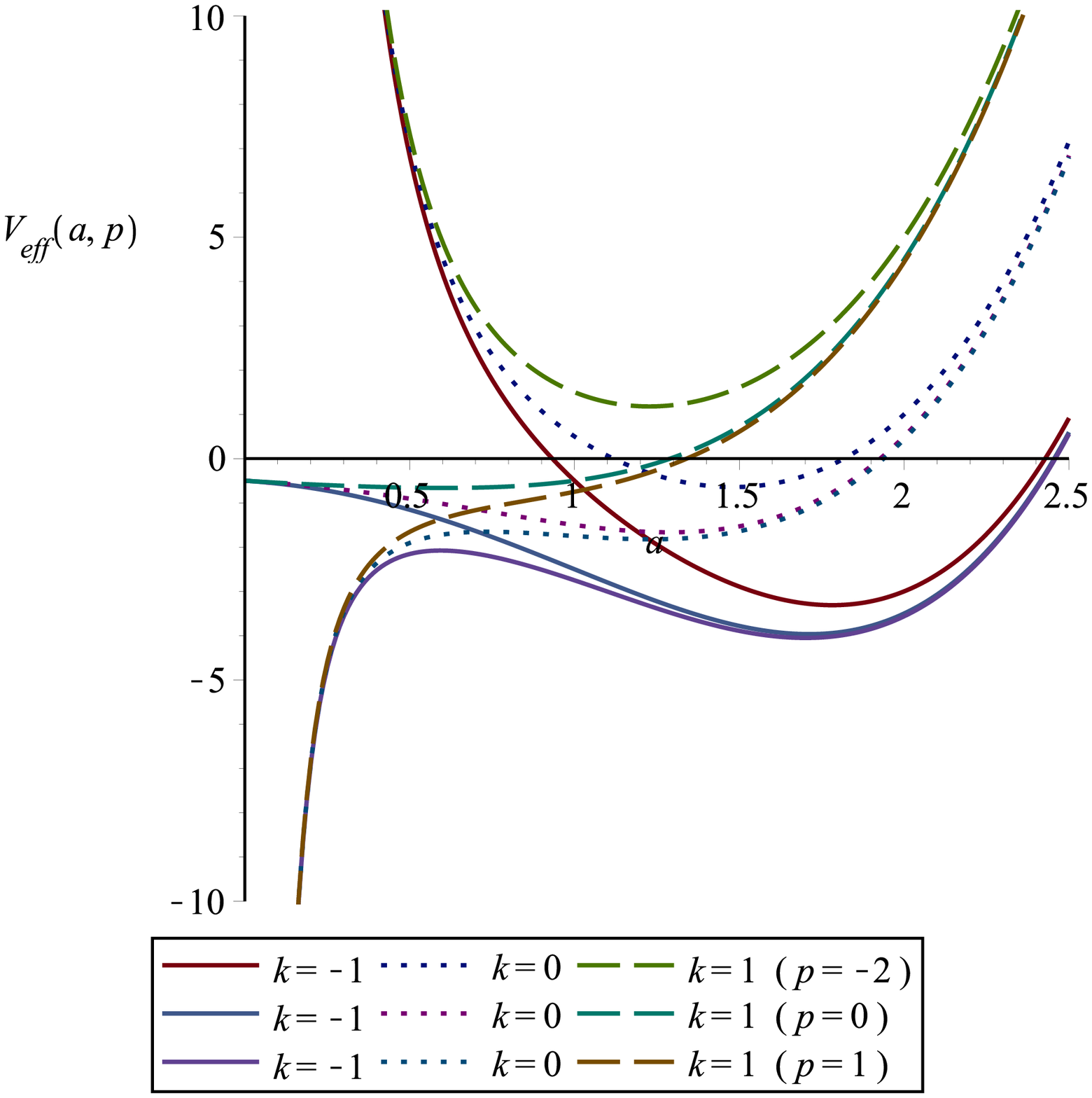}
	\caption{The effective potential $V_{eff}(a,p)$ for $\Lambda = -|\Lambda|$. We focus on the $p=-2,1$ cases and compare with the case where $p=0$.}
	\label{fig:WDW_Fig5}
\end{figure}

\begin{figure}[!ht]
		\includegraphics[scale=0.35]{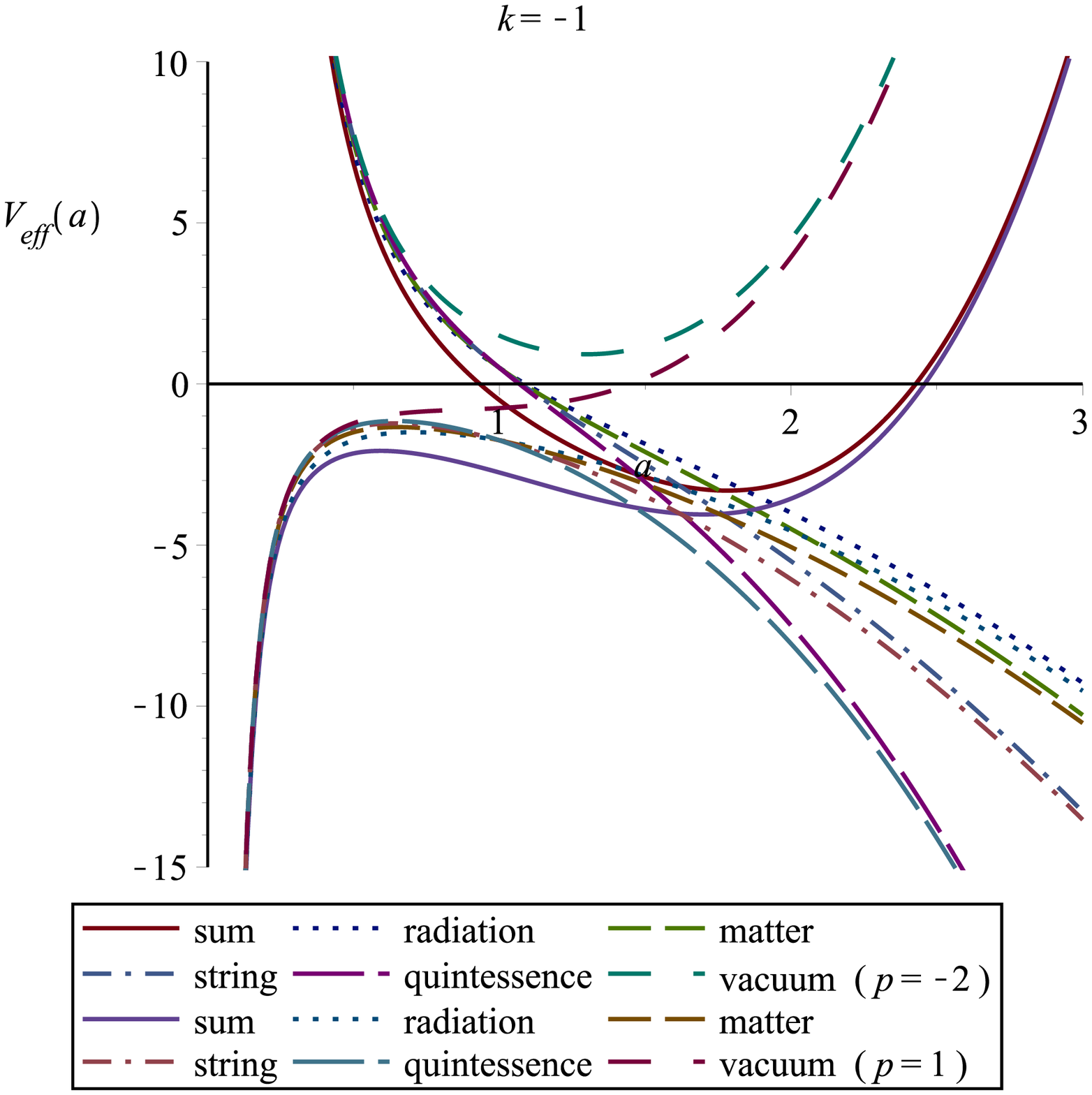}
	\caption{The effective potential $V_{eff}(a,p)$ for $\Lambda = -|\Lambda|$.  We focus on the $k=-1$ case and compare with each kind of energy density $\omega$.}
	\label{fig:WDW_Fig6}
\end{figure}

\begin{figure}[!ht]
		\includegraphics[scale=0.35]{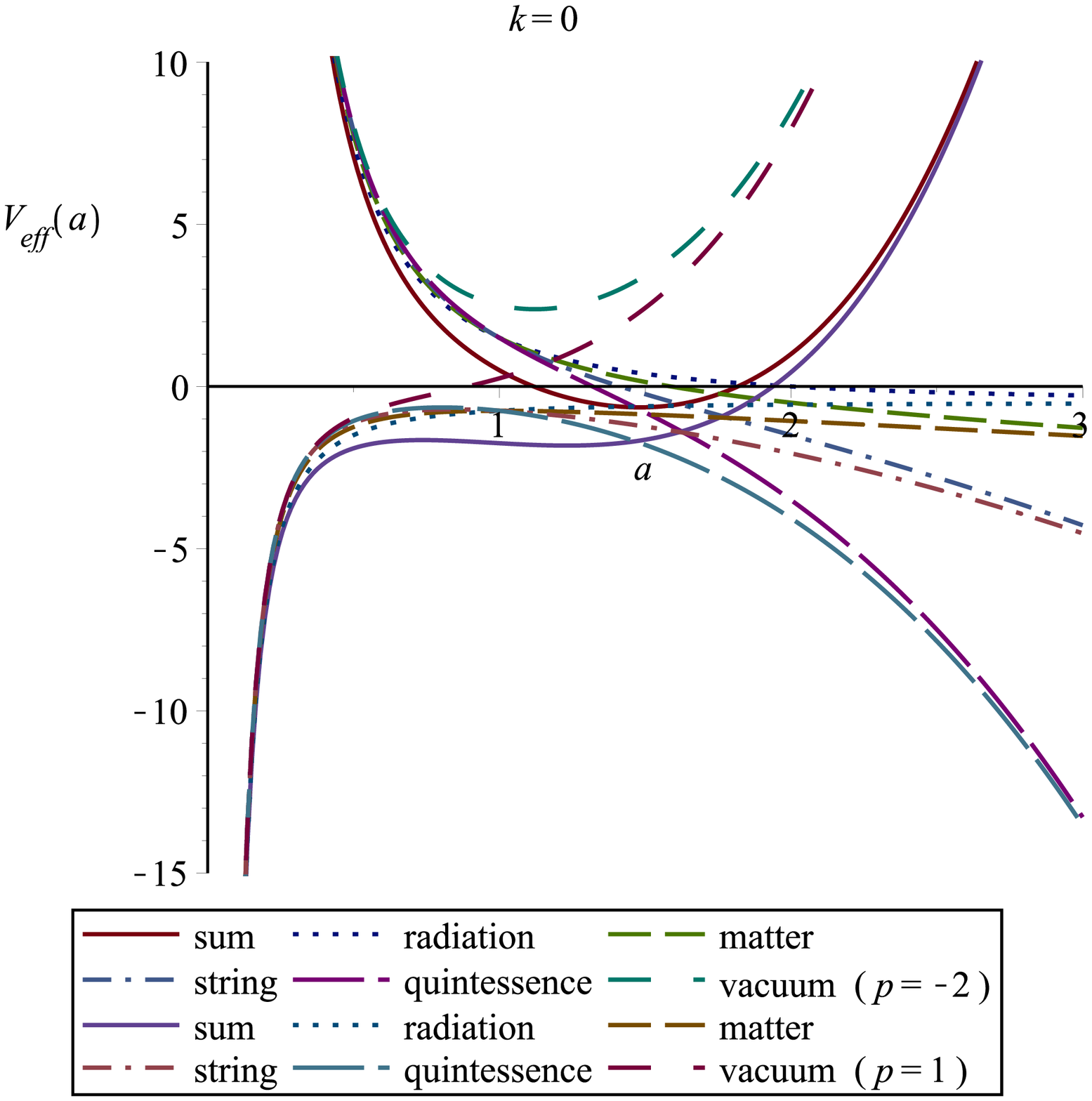}
	\caption{The effective potential $V_{eff}(a,p)$ for $\Lambda = -|\Lambda|$.  We focus on the $k=0$ case and compare with each kind of energy density $\omega$.}
	\label{fig:WDW_Fig7}
\end{figure}

\begin{figure}[!ht]
		\includegraphics[scale=0.35]{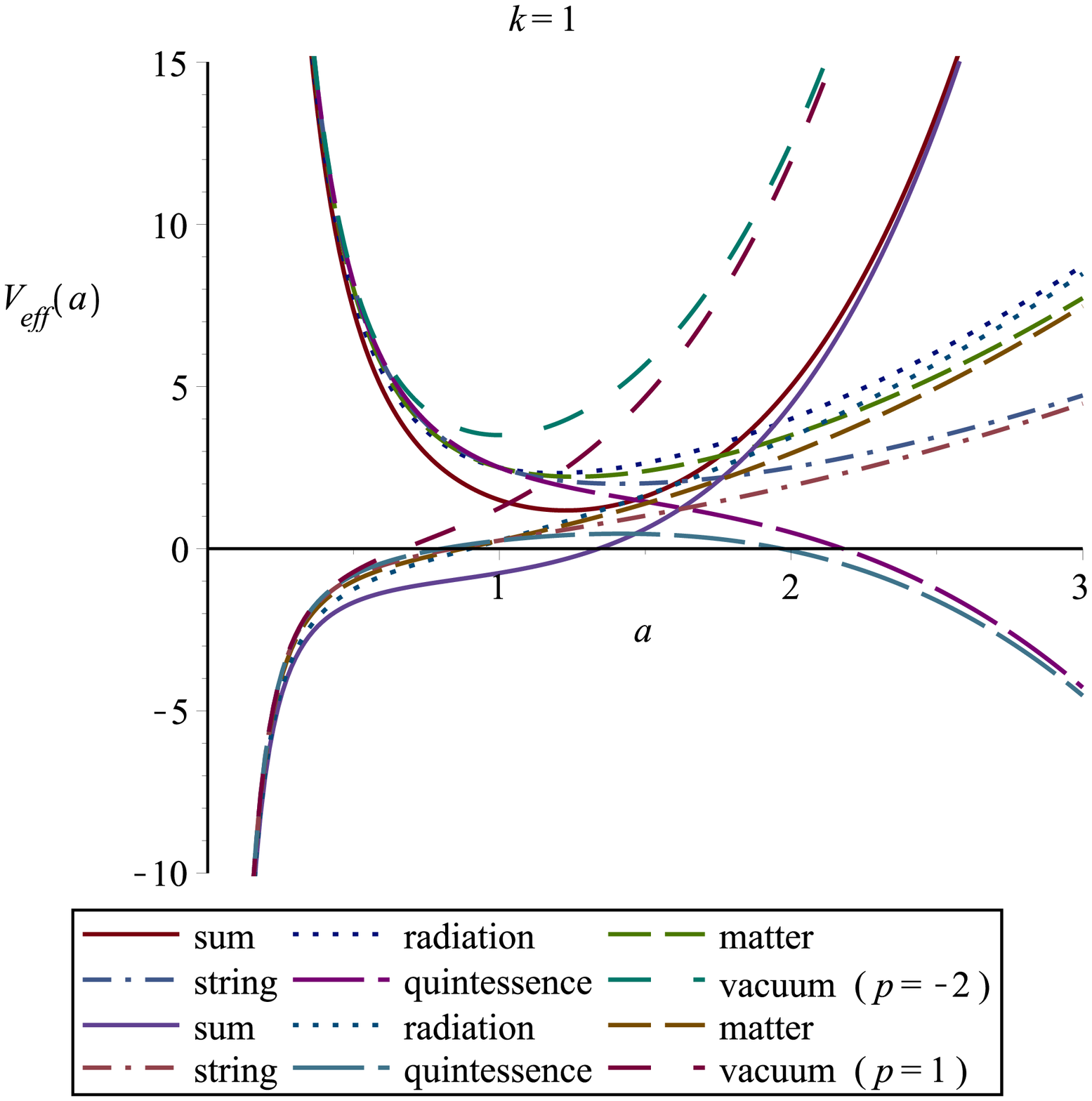}
	\caption{The effective potential $V_{eff}(a,p)$ for $\Lambda = -|\Lambda|$.  We focus on the $k=+1$ case and compare with each kind of energy density $\omega$.}
	\label{fig:WDW_Fig8}
\end{figure}

Thus, we will discuss the effective potential of the Wheeler-DeWitt equation for these two values of the ordering parameter and compare with the case when $p=0$ (or $p=2$), in which case we recover the results known in the literature. The behaviors of $V_{eff}(a,p)$ with $p=-2,1$ are shown in Figures~\ref{fig:WDW_Fig1}-\ref{fig:WDW_Fig4} and \ref{fig:WDW_Fig5}-\ref{fig:WDW_Fig8}, for positive and negative values of the cosmological constant, respectively.

In Figure~\ref{fig:WDW_Fig1} we see that the effective potential for a positive value of the cosmological constant is finite near the origin only in the case when $p=0$ (or $p=2$), while the cases when $p=-2$ and $p=1$ provides asymptotic values for $V_{eff}(a,p)$. The same can be concluded from Figure~\ref{fig:WDW_Fig5}, with a negative value of the cosmological constant. Therefore, it seems that some boundary condition may exist for $p=0$ and $p=2$. Both figures show us that there is a strong condition between $V_{eff}(a,p)$ and the values of the ordering parameter, quantitatively and qualitatively as well.

From Figures~\ref{fig:WDW_Fig2}-\ref{fig:WDW_Fig4} we conclude that the behavior of the effective potential for a positive value of the cosmological constant near the origin and at the end of the expansion is dominated by the energy density due to the radiation and vacuum, respectively. The same occurs in Figures~\ref{fig:WDW_Fig6}-\ref{fig:WDW_Fig8}, for a negative value of the cosmological constant. Therefore, we see that these two forms of energy density will play a very important role on the wave function of the universe. It is worth calling attention to the fact that the behavior of the effective potential shown in Figures~\ref{fig:WDW_Fig2}-\ref{fig:WDW_Fig4}, for positive values of the cosmological constant, as well as for its negative values, as shown in Figures~\ref{fig:WDW_Fig6}-\ref{fig:WDW_Fig8}, are only slightly influenced by the geometry of the spatial section and strongly influenced by the ordering parameter.

Thus, from Figures~\ref{fig:WDW_Fig1}-\ref{fig:WDW_Fig8}, we can conclude that the term which involves the ordering parameter $p$ will contribute in a significant way, only at the early stages of the universe, in other words when $a \rightarrow 0$.

Now, let us solve Eq.~(\ref{eq:WDE_mov_1}) in a general way, in a sense that the solution will be valid for all forms of energy density. In order to do this, we can write the Wheeler-DeWitt equation with an arbitrary ordering parameter in the Friedmann-Robertson-Walker universe as
\begin{eqnarray}
&& \frac{d^{2}y(a)}{da^{2}}+\biggl(B_{0}+B_{1}a+B_{2}a^{2}+B_{3}a^{3}\nonumber\\
&& +B_{4}a^{4}+\frac{B_{5}}{a^{2}}\biggr)y(a)=0,
\label{eq:generalized_WDW_equation_ordering_FRW_universe}
\end{eqnarray}
where the coefficients $B_{0}$, $B_{1}$, $B_{2}$, $B_{3}$, $B_{4}$, and $B_{5}$ are given by
\begin{equation}
B_{0}=\frac{6\pi^{3}c^{2}}{\hbar^{2}G}A_{r},
\label{eq:B0_FRW_universe}
\end{equation}
\begin{equation}
B_{1}=\frac{6\pi^{3}c^{2}}{\hbar^{2}G}A_{m},
\label{eq:B1_FRW_universe}
\end{equation}
\begin{equation}
B_{2}=\frac{6\pi^{3}c^{2}}{\hbar^{2}G}A_{s}-\frac{9\pi^{2} c^{6}}{4\hbar^{2}G^{2}}k,
\label{eq:B2_FRW_universe}
\end{equation}
\begin{equation}
B_{3}=\frac{6\pi^{3}c^{2}}{\hbar^{2}G}A_{q},
\label{eq:B3_FRW_universe}
\end{equation}
\begin{equation}
B_{4}=\frac{6\pi^{3}c^{6}}{8\pi\hbar^{2}G^{2}}\Lambda,
\label{eq:B4_FRW_universe}
\end{equation}
\begin{equation}
B_{5}=\frac{(2 p-p^2)}{4}.
\label{eq:B5_FRW_universe}
\end{equation}
No general solution in terms of standard functions is known for this equation over the entire range $0 \leq a < \infty$. However, we may obtain solutions near $a = 0$ and for $a \rightarrow \infty$, as follows.
%
%
\subsection{Case 1: scale factor very small (\texorpdfstring{$a \rightarrow 0$}{a->0})}
By expanding all terms in Eq.~(\ref{eq:generalized_WDW_equation_ordering_FRW_universe}) we have for small $a$, which means when the universe was very small, the following equation
\begin{equation}
\frac{d^{2}y(a)}{da^{2}}+\biggl(B_{0}+\frac{B_{5}}{a^{2}}\biggr)y(a)=0.
\label{eq:WDW_equation_ordering_FRW_universe_case1}
\end{equation}
Therefore, we can write the analytical solutions of Eq.~(\ref{eq:WDW_equation_ordering_FRW_universe_case1}) in terms of the Bessel functions as
\begin{equation}
y(a)=a^{\frac{1}{2}}[C_{1}\ i\ J_{\nu}(\sqrt{B_{0}}a)+C_{2}\ N_{\nu}(\sqrt{B_{0}}a)],
\label{eq:y_Bessel_WDW_equation_ordering_FRW_universe}
\end{equation}
where $J_{\nu}$ and $N_{\nu}$ are the Bessel functions of the first and second (Neumann functions) kind, respectively, with $C_{1}$ and $C_{2}$ being constants to be determined. The parameter $\nu$ is defined by
\begin{equation}
\nu=\frac{\sqrt{1-4B_{5}}}{2}=\frac{p-1}{2}.
\label{eq:parameter_nu_WdW_ordering}
\end{equation}

Taking into account the series expansion of the Bessel functions, for small $x$, given by \cite{Arfken:2005}
\begin{equation}
J_{\nu}(x) \sim \frac{1}{\nu!}\biggl(\frac{x}{2}\biggr)^{\nu}+\ldots,
\label{eq:BesselJ_zero_WdW_ordering}
\end{equation}
\begin{equation}
N_{\nu}(x) \sim -\frac{(\nu-1)!}{\pi}\biggl(\frac{x}{2}\biggr)^{-\nu}+\ldots,
\label{eq:BesselN_zero_WdW_ordering}
\end{equation}
we obtain that when $a \rightarrow 0$ the solution given by Eq.~(\ref{eq:y_Bessel_WDW_equation_ordering_FRW_universe}) can be written as
\begin{eqnarray}
\Psi(a) & \sim & C_{1}\ i\ \frac{1}{\nu!}\biggl(\frac{\sqrt{B_{0}}}{2}\biggr)^{\nu}a^{\frac{p}{2}}\nonumber\\
&& -C_{2}\ \frac{(\nu-1)!}{\pi}\biggl(\frac{\sqrt{B_{0}}}{2}\biggr)^{-\nu}a^{\frac{2-p}{2}}.
\label{eq:Psi_zero_WdW_ordering}
\end{eqnarray}

From this solution we can conclude that the wave function approaches a constant when $a \rightarrow 0$, for $0 \leq p \leq 2$. In this case $C_{1}$ and $C_{2}$ can assume any value without violating what $\Psi(a)$ means. In the case $p \leq -1$ or $p \geq 3$, $\Psi(a)$ will be divergent, but in order to avoid this behavior, which is incompatible with the meaning of the wave function, we must impose $C_{1}=0$ or $C_{2}=0$, respectively, and hence we have that $\Psi(a \rightarrow 0) \sim 0$.

Therefore, from a dynamical interpretation of the wave function of the universe, we have obtained a boundary condition for the ordering parameter, namely, $0 \leq p \leq 2$. Note that when $a \rightarrow 0$, for $p=0$ and $p=2$, we have to determine the constants $C_{1}$ and $C_{2}$, respectively, so that the solution satisfies a boundary condition, as for example, the wave function $\Psi(a)$ should be normalized to unity for any $a$ (very small). Finally, for $p=1$, we have that $\Psi(a \rightarrow 0) \sim 0$, and hence it is not necessary to fix any constant.

It is worth calling attention to the fact that this asymptotic solution is valid for any value of the curvature parameter ($k=-1,0,+1$), which means that does not depend on the geometry of the spatial sector. It depends on the energy density of the radiation (coefficient $B_{0}$), which is the predominant form of energy at early stages of the universe. From this point of view, we are extending the solution obtained by He \textit{et al.} \cite{PhysLettB.748.361}, which is valid only for the case of the flat universe $k=0$.

\begin{figure}[!t]
		\includegraphics[scale=0.90]{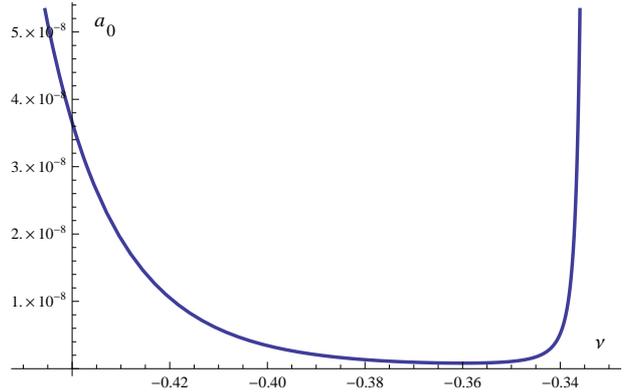}
	\caption{Minimum scale factor as a function of $\nu$, for $B_0=1$ and $C=0.008$.}
	\label{fig:WDW_Fig9}
\end{figure}

We can also obtain a relationship between the quantity associated to the radiation energy density of the universe in a specific instant, $B_0$, and a postulated initial value for the scale factor, $a_0$, by considering the maximum probability density at such a value, calculated from $\partial_a |\Psi(a)|^2=0$ \cite{PhysDarkUniv.28.100547}. Taking into account Eq.~(\ref{eq:y_Bessel_WDW_equation_ordering_FRW_universe}), the value of $a_0$ arises thus from the leading terms of the derivative expansion,
\begin{eqnarray}
a_0 & \approx & \frac{\pi ^{-2/\nu}}{\sqrt{B_0}} \bigg\{\frac{2^{\nu+2} \cos (\pi  \nu) \Gamma (\nu+1)^2}{C^2 [\Gamma (\nu)+4 \Gamma (\nu+1)]}\bigg\}^{1/\nu}\nonumber\\
&& \times [2 \Gamma (-\nu) \Gamma (\nu+1)+\Gamma (-\nu) \Gamma (\nu)\nonumber\\
&& -\Gamma (1-\nu) \Gamma (\nu)]^{1/\nu},
\end{eqnarray}
in the limit for which $a\ll 1$, where we have chosen $C_{1}=C_{2}=C\sqrt{\pi/2}$. In Figure~(\ref{fig:WDW_Fig9}) we depict the minimum scale factor, which corresponds to the maximum probability density near the origin, as a function of the parameter $\nu$. Notice the local minimum around $\nu \approx -0.36$.

%
%
\subsection{Case 2: large scale factor (\texorpdfstring{$a \rightarrow \infty$}{a->infinity})}
We now expand all terms in Eq.~(\ref{eq:generalized_WDW_equation_ordering_FRW_universe}) for large $a$, which means the scale factor at the end of the expansion, and write the equation in the form
\begin{eqnarray}
&& \frac{d^{2}y(a)}{da^{2}}+\biggl(B_{0}+B_{1}a+B_{2}a^{2}+B_{3}a^{3}\nonumber\\
&& +B_{4}a^{4}\biggr)y(a)=0.
\label{eq:WDW_equation_ordering_FRW_universe_case2}
\end{eqnarray}
Therefore, we can write the analytical solution of Eq.~(\ref{eq:WDW_equation_ordering_FRW_universe_case2}) in terms of the Triconfluent Heun function. It is given by
\begin{eqnarray}
y(x) & = & \mbox{e}^{-\frac{1}{2}(x^{3}+\gamma x)}\nonumber\\
&& \times \{C_{1}\ \mbox{HeunT}(\alpha,\beta,\gamma;x)\nonumber\\
&& +C_{2}\ \mbox{e}^{x^{3}+\gamma x}\ \mbox{HeunT}\biggl(\lambda^{2}\alpha,-\beta,-\lambda\gamma;\frac{x}{\lambda}\biggr),\nonumber\\
\label{eq:y_HeunT_WDW_equation_ordering_FRW_universe}
\end{eqnarray}
where $\lambda^{3}=-1$. The new variable $x$ is defined by
\begin{equation}
x=\tau(a+\xi),
\label{eq:x_FRW_universe}
\end{equation}
where the parameters $\tau$ and $\xi$ are expressed as
\begin{equation}
\tau=\biggl(-\frac{4B_{4}}{9}\biggr)^{1/6},
\label{eq:tau_FRW_universe}
\end{equation}
\begin{equation}
\xi=\frac{B_{3}}{4B_{4}}.
\label{eq:xi_FRW_universe}
\end{equation}
The parameters $\alpha$, $\beta$, and $\gamma$ are defined by
\begin{equation}
\alpha=b_{0}+\frac{1}{9}b_{2}^{2},
\label{eq:alpha_FRW_universe}
\end{equation}
\begin{equation}
\beta=b_{1},
\label{eq:beta_FRW_universe}
\end{equation}
\begin{equation}
\gamma=-\frac{2}{3}b_{2},
\label{eq:gamma_FRW_universe}
\end{equation}
where the coefficients $b_{0}$, $b_{1}$, and $b_{2}$ are identified by
\begin{equation}
b_{0}=\frac{B_{0}-B_{1} \xi +B_{2} \xi ^2-B_{3} \xi ^3+B_{4} \xi ^4}{\tau ^2},
\label{eq:b0_FRW_universe}
\end{equation}
\begin{equation}
b_{1}=\frac{B_{1}-2 B_{2} \xi +3 B_{3} \xi ^2-4 B_{4} \xi ^3}{\tau ^3},
\label{eq:b1_FRW_universe}
\end{equation}
\begin{equation}
b_{2}=\frac{B_{2}-3 B_{3} \xi +6 B_{4} \xi ^2}{\tau ^4}.
\label{eq:b2_FRW_universe}
\end{equation}

On the other hand, the series expansion of the triconfluent Heun functions, for large $x$, are given by \cite{Ronveaux:1995}
\begin{eqnarray}
\mbox{HeunT}(\alpha,\beta,\gamma;x) & \sim & C_{1}\ x^{\frac{\beta}{3}-1}\nonumber\\
&& \times \sum_{n \geq 0}a_{n}(\alpha,\beta,\gamma)x^{-n}\nonumber\\
&& + C_{2}\ x^{-\frac{\beta}{3}-1}\ \mbox{e}^{x^{3}+\gamma x}\nonumber\\
&& \times \sum_{n \geq 0}(-1)^{n}a_{n}(\alpha,-\beta,\gamma)x^{-n},\nonumber\\
\label{eq:HeunB_infty_FRW_universe}
\end{eqnarray}
where $|\arg x| \leq \pi/2$, and $a_{0}(\alpha,\pm\beta,\gamma)=1$. Thus, when $a \rightarrow \infty$ implies that $x \rightarrow \infty$, we have the following asymptotic solution
\begin{equation}
\Psi(x) \sim C\ \biggl(\frac{x}{\tau}-\xi\biggr)^{-\frac{p}{2}}\frac{1}{x}\cosh\biggl[\frac{\beta}{3}\ln x-\frac{1}{2}(x^{3}+\gamma x)\biggr].
\label{eq:Psi_infinity_WdW_ordering}
\end{equation}
Here, for simplicity and convenience, we have chosen the free parameters $C_{1}$ and $C_{2}$ equal to $C/2$, where $C$ is a constant to be determined using the condition that $\Psi(x)$, given by Eq.~(\ref{eq:Psi_infinity_WdW_ordering}), should be appropriately normalized. As it should be expected, the values of $C_{1}$ and $C_{2}$ in this regime, namely, $a \rightarrow \infty$, should be different from that which occur for $a \rightarrow 0$.

It is worth mention that in this case, the wave function of the universe goes to zero when $a \rightarrow \infty$, independently of the value of the ordering parameter. Therefore, as expected from a dynamical point of view, the ordering term does not play a role in the behavior of the universe at the end of the expansion.
%
%
\section{Summary}\label{Summary}
This work has illustrated the difficulty of solving the Wheeler-DeWitt equation with an ordering parameter in the Friedmann-Robertson-Walker universe when it is taken into account several kinds of energy and considering any value of curvature. In the absence of analytical solutions, we examined the solutions only for very small ($a \rightarrow 0$) and very large values of the scale factor ($a \rightarrow \infty$).

We have found the asymptotic behavior of the wave function of the universe in a quantum cosmological scenario. In the Cases 1 and 2 therefore, the asymptotic solutions of Eq.~(\ref{eq:generalized_WDW_equation_ordering_FRW_universe}) near the origin and at infinity are expressed in terms of Bessel and triconfluent Heun functions, respectively. From these results, we obtained a boundary condition for the ordering parameter, namely, $0 \leq p \leq 2$, which implies $-0.5 \leq \nu \leq 0.5$.

Supposing that the universe has started its existence with a minimum size, we have obtained the corresponding minimum scale factor, $a_0$, which is related to the maximum probability density near the origin, $a=0$, which is expressed as a function of the radiation density energy and of the parameter $\nu=(p-1)/2\leq 0$. Though the probabilistic interpretation of the universe wave function is still controversial, it is admitted by some authors \cite{PhysLettB.748.361}. Furthermore, we have shown that the ordering term does not contribute in an significant way at the end of the expansion of the universe. Therefore, the ordering parameter has only a quantum dynamical interpretation at the early stages of the universe.

Finally, the approach developed here leads to a full representation of the solutions at the asymptotic regimes in terms of standard functions in the most general case. It is worth calling attention to the fact that for many quantum calculations such solutions are all that should be required.
%
%
\section*{Acknowledgement}
H.S.V. is funded by the Coordena\c c\~{a}o de A\-per\-fei\-\c co\-a\-men\-to de Pessoal de N\'{i}vel Superior - Brasil (CAPES) - Finance Code 001. V.B.B. is partially supported from CNPq Project No. 305835/2016-5. M.S.C. is partially supported from CNPq Project Numbers 433168/2016-1 and 314183/2018-3. The authors would like to thank FUNCAP for partial financial support under the grant PRONEM PNE-0112-00085.01.00/16. The authors also would like to thank Professor Daniel Vanzella for the fruitful discussions.
%
%

%
%
\end{document}